\documentstyle[12pt]{article}

\begin{document}

\begin{titlepage}

\begin{flushright}
DFPD-01/TH/12
\end{flushright}

\vspace{1cm}
\begin{center}
{\Large \bf Double chiral logarithms of Generalized Chiral Perturbation Theory
for low-energy $\pi\pi$ scattering}

\vskip .5in 

{\normalsize \bf Luca Girlanda

\vskip .5in
 
{\small \it Dipartimento di Fisica ``Galileo Galilei'', Universit\`a di
Padova and INFN \\ Via Marzolo 8, I-35131 Padova, Italy}   
    }

\vskip .5in

\begin{abstract}

{\normalsize
We express the two-massless-flavor Gell-Mann--Oakes--Renner ratio in terms of
low-energy $\pi\pi$ observables including the $O(p^6)$ double chiral
logarithms of generalized chiral perturbation theory.
Their contribution is sizeable and tends to compensate the one from the single
chiral logarithms. However it is not large enough to spoil the convergence
of the chiral expansion. As a signal of reduced theoretical uncertainty, we
find that the scale dependence from the one-loop
single logarithms is almost completely canceled by the one from the two-loop
double logarithms.
}

\end{abstract}

\vskip 1in

\end{center}

\vfill


\vfill

\end{titlepage}

\indent

{\bf 1.}~ The dynamics of chiral symmetry breakdown ($\chi$SB) in the
Standard Model
remains an open issue both from the experimental and theoretical standpoint.
At very low energy the problem concerns the understanding of the mechanism
 of $\chi$SB for QCD. In this framework it is generally assumed that chiral
symmetry is broken through the formation of a quark-antiquark condensate.
A theoretically consistent
alternative is that the symmetry is broken by higher dimensional order
parameters, analogously to what happens in antiferromagnetic spin systems
for the rotational symmetry \cite{stern}. Deeper insight into this problem
can be
gained by considering the theory in a Euclidean box, with (anti)periodic
boundary conditions. The two alternatives are then distinguished by a
qualitatively different behavior of the smallest eigenvalues of the Dirac
operator in the thermodynamical limit.
In fact it has recently been realized that a prominent role is played by the
number of massless flavors $N_f$ of the theory \cite{nfdep,para}. In general, as
$N_f$ increases, the order parameters dominated by the infrared end of the
spectrum of the Euclidean Dirac operator experience a paramagnetic
suppression which, for  $N_f$ large enough, will eventually restore chiral
symmetry. Since only the SU(2) and SU(3) chiral limits are reasonable
approximations to the real world, the question to be addressed is to which
extent $N_f=2,3$ is close to a critical point in which $\langle \bar q q
\rangle$ disappears and chiral symmetry is eventually broken by
higher-dimensional order parameters. 
 As was pointed out in Refs.~\cite{pipi0,gchpt}, it is possible to submit the
standard hypothesis of a large condensate to experimental verification, through the measurement of
low-energy $\pi\pi$ phase-shifts. 
 The S-wave $\pi\pi$ interaction at low energy is very sensitive to the
two-massless-flavor chiral condensate: it  gets stronger as the condensate
decreases. Standard chiral perturbation theory (ChPT) \cite{gale} pushed to two-loop accuracy,
together with the use of constraints from Roy equations \cite{roy}, provides a very
sharp prediction for the S-wave scattering lengths \cite{sta0a2}. 
The situation has
evolved recently, since new preliminary experimental data provided by the E865
experiment on $K_{e4}$ decays have become available \cite{truol}.
In fact
a first analysis of the above mentioned data in the standard context indicates that
deviations from this scenario are marginal \cite{kl4con}.
On the other hand, generalized chiral perturbation theory (GChPT), which relaxes from
the beginning the standard assumption, cannot produce any prediction for the
low-energy $\pi\pi$ observables, but only relates these observables to the
magnitude of the quark condensate.
It is the purpose of the present work to establish the relationship between 
 low-energy $\pi\pi$ observables and the two-massless-flavors quark
condensate in 
the language of GChPT beyond the $O(p^4)$ level, i.e. including the $O(p^6)$
double chiral logarithms, which are among the potentially most dangerous
two-loop contributions.

\indent
%\section{The GOR ratio}

{\bf 2.}~ Relaxing the standard assumption of a dominant condensate implies
that the Gell-Mann--Oakes--Renner ratio\footnote{ In Eq.~(\ref{eq:gor}),
$F_\pi^2 M_\pi^2$ should be considered simply as physical units for 
measuring the QCD renormalization group invariant combination $(m_u + m_d)
\langle \bar q q \rangle$.},
\begin{equation} \label{eq:gor}
x_2^{\mathrm{GOR}}= \frac{(m_u + m_d) }{F_{\pi}^2
M_{\pi}^2} \lim_{m_u,m_d \to 0} | \langle \bar q q\rangle |,
\end{equation}
is actually a free parameter. If its value were not close to 1, then the
quadratic term in quark masses would be as important as the condensate term
in the chiral expansion of the squared pion mass. Thus
the counting rule has to be 
modified accordingly: both the light quark mass $\hat m$ ($m_u=m_d=\hat
m$) and the quark condensate are quantities of chiral order $O(p)$.
The effective Lagrangian is organized as an infinite sum ${\mathcal L}= {\mathcal L}^{(2)} + {\mathcal
L}^{(3)} + {\mathcal L}^{(4)} + \ldots $ where the $O(p^d)$ term ${\mathcal
L}^{(d)}$ contains terms with $k$ derivatives, $l$ powers of quark masses
and $n$ powers of the quark condensate, such that $k+l+n=d$. 
It contains, at each finite chiral order, additional operators with respect
to the corresponding standard Lagrangian.
For instance, the leading $O(p^2)$ Lagrangian reads,
\begin{eqnarray} \label{eq:p2lagr}
{\mathcal L}^{(2)} &=&
\frac{F^2}{4} \left\{ \langle D_{\mu} U^{\dagger} D^{\mu} U \rangle + 2 B
\langle U^{\dagger} \chi + \chi^{\dagger} U \rangle + A \langle \left(
U^{\dagger} \chi \right)^2 + \left( \chi^{\dagger} U \right)^2 \rangle
\right. \nonumber \\
&& \left. + Z^P \langle U^{\dagger} \chi - \chi^{\dagger} U\rangle^2 + h_0
\langle \chi^{\dagger} \chi \rangle + h_0^{\prime} \left( \det \chi + \det
\chi^{\dagger} \right) \right\},
\end{eqnarray}
where $U \in$~SU(2) collects the Goldstone bosons' degrees of freedom and
$\chi$ denotes the scalar-pseudoscalar external source, to be expanded
around the quark mass matrix $\chi = \hat m {\bf 1} +
\ldots $. Notice that, compared to standard notations, a factor $2 B$ has
been removed from the definition of $\chi$, for consistency with the new
chiral counting: in fact $B$ is related to the quark condensate, $F^2 B = -
\lim_{\hat m \to 0} \langle \bar q q \rangle$, and it has to be considered,
formally, as a small parameter, $B \sim \chi \sim O(p)$.
One-loop diagram with insertions from the Lagrangian~(\ref{eq:p2lagr}) are
of chiral order $O(p^4)$. They are divergent and require the renormalization
of the low-energy constants (l.e.c.). In the minimal subtraction scheme of
dimensional regularization,
\begin{equation} \label{eq:lecren}
{\mathrm{const}}={\mathrm{const}}^r + 
\frac{\mu^{d-4}}{16 \pi^2} \frac{\Gamma_{\mathrm{const}}}{d-4},
\end{equation}
where $\mu$ is the renormalization scale.
The $\beta$-function coefficients $\Gamma_{\mathrm{const}}$ for the leading
l.e.c.'s are 
\begin{equation} \label{eq:p2ren}
F^2 \Gamma_A = 3 B^2, \quad F^2 \Gamma_{Z^P} = -\frac{3}{2} B^2, \quad F^2
\Gamma_{h_0^\prime} = 6 B^2,
\end{equation}
whereas $F^2$, $B$ and $h_0$ do not get renormalized.
The renormalized constants are renormalization scale dependent, according to
\begin{equation} \label{eq:lecbeta}
\mu \frac{d}{d \mu} {\mathrm{const}}^r =- \frac{1}{16 \pi^2}
\Gamma_{\mathrm{const}}.
\end{equation}
Due to the chiral counting of the condensate parameter $B$, the divergences
arising from the renormalization of the l.e.c.'s are of chiral order $p^4$,
as they should, in order to absorb the ones arising from the loops.
The renormalization of the leading l.e.c.'s~(\ref{eq:p2ren}) is not
sufficient to absorb all the one-loop $O(p^4)$ divergences generated by
${\mathcal L}^{(2)}$, which is a consequence of the fact that the
Lagrangian~(\ref{eq:p2lagr}) is non-renormalizable. 
One has to include also the subleading $O(p^3)$ and $O(p^4)$ Lagrangians and
renormalize the relative l.e.c.'s, according to Eq.~(\ref{eq:lecren}).
The $\beta$-function coefficients $\Gamma_{{\mathcal L}^{(3)}}$ for the
l.e.c.'s of  ${\mathcal L}^{(3)}$ will be proportional to $B$, in order that
the divergences be of order $p^4$, while those for the l.e.c.'s of
${\mathcal L}^{(4)}$ will be independent of $B$,
\begin{equation}
\Gamma_{{\mathcal L}^{(2)}} \sim O(B^2) \sim O(p^2), \quad \Gamma_{{\mathcal
L}^{(3)}} \sim O(B) \sim O(p), \quad \Gamma_{{\mathcal L}^{(4)}} \sim O(1).
\end{equation}
We refer to existing literature for the complete list of operators together
with their renormalization up to $O(p^4)$ \cite{su2lagr}. 

\indent 

\begin{figure}[ht] 
\vspace{10cm}
\includegraphics{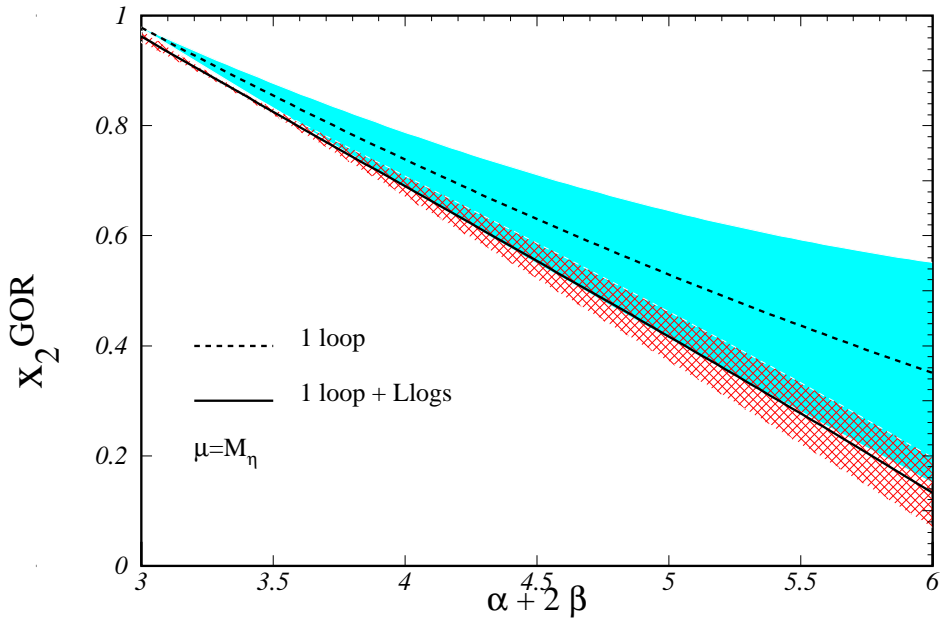}
 \caption{\it The GOR ratio at 1 loop including (continuous) and not
including (dashed) the double logarithms. The scale is set to $\mu=M_\eta
\pm 250$~MeV.   
    \label{fig:xgor} }
\end{figure}

{\bf 3.}~ The proliferation of chirally invariant operators has no bearing on the form
of the $\pi\pi$ scattering amplitude up to two-loop level: the latter
only depends on
six ``observable'' parameters, independently of the chiral counting.
This is a consequence of unitarity, analyticity and crossing symmetry,
and of the chiral suppression of partial waves greater than two \cite{pipi0}.
Neglecting $O(p^8)$, i.e. at two-loop level, the invariant amplitude for
$\pi\pi$ scattering has been given,
using dispersive methods, in Ref.~\cite{pipi1} in 
terms of the parameters $\alpha,\beta,\lambda_1,\ldots ,\lambda_4$,
\begin{equation} \label{eq:akmsf}
 A(s|t,u) = A_{\mathrm{KMSF}} (s|t,u;\alpha,\beta;\lambda_1, \lambda_2, \lambda_3,
  \lambda_4) 
% + O \left[ \left( \frac{p}{ \Lambda_{\mathrm{H}}} \right)^8 , \left(
%\frac{M_{\pi}}{
% \Lambda_{\mathrm{H}}} \right)^8 \right].
+ O(p^8).
\end{equation}
The parameter $\alpha$ represents, at leading order, the amplitude at the
symmetric point $s=t=u=4/3 M_\pi^2$, and $\beta$ its slope. These two
parameters can be viewed as the subtraction constants in a Roy-like dispersive
representation of the amplitude which fixes, confronted to existing
experimental data in the medium and high energy region, the remaining 4
parameters $\lambda_1,\ldots ,\lambda_4$ \cite{pipi2}.
A brute force calculation in ChPT allows one to find the ``chiral anatomy''
of these parameters, i.e.\, their expressions in terms of quark masses
and l.e.c.'s.
In particular we are interested in finding their dependence on the
l.e.c.~$B$, related to the quark condensate.
At tree level only the parameter $\alpha$  depends on the condensate: its
value varies from 1 to 4 if $\langle \bar q q \rangle$ decreases from its
standard value down to zero, while $\beta$ is equal to~1 and the
$\lambda_i$'s vanish.
Going to one-loop precision already involves  a conspicuous set of l.e.c.
characteristics of GChPT, besides the contribution of the chiral
logarithms.
Explicit expressions for the  1-loop parameters $\alpha$, $\beta$,
$\lambda_1$ and
$\lambda_2$ can be found in Refs.~\cite{su2lagr}.
After eliminating as many unknown constants as possible in favor of physical
quantities, one can invert these
expressions for the quark condensate, and  
write the GOR ratio in terms of the combination $\alpha + 2 \beta$, modulo
remaining unknown constants, renormalized at a scale $\mu$ \cite{monpi}:
\begin{eqnarray} \label{eq:xgor1l}
x_2^{\mathrm{GOR}} & = &  \frac{2 \hat m B F^2}{F_{\pi}^2 M_{\pi}^2} = 2 -
\frac{ \alpha + 2 \beta}{3} + \frac{F^2}{F_{\pi}^2 M_{\pi}^2} \left( 15
\rho_1^{r} - \rho_2^{r} + 28 \rho_4^{r} - 2 \rho_5^{r} \right) \hat
m^3  \nonumber \\
&& + \left[ 4 a_2^r +   8 \left( \frac{\alpha + 2 \beta }{3} - 1  \right) a_3
+ 8 b_2^r + 16 c_1^r \right]
\hat m^2 \nonumber \\
&& + \frac{64}{M_{\pi}^2} ( e_1^r + 2 f_1^r + 2 f_2^r + 4 f_4^r ) \hat m^4
 \nonumber \\
 && + \frac{M_{\pi}^2}{288 \pi^2 F_{\pi}^2} \left(
\alpha + 2 \beta \right) \left[ 24 - 11 ( \alpha + 2 \beta) \right]
\nonumber \\
&& +
 \left[ 6 + \frac{5}{3} (\alpha + 2 \beta) - \frac{11}{9} (\alpha + 2
 \beta)^2 \right] \frac{M_{\pi}^2}{32 \pi^2 F_{\pi}^2} \log
 \frac{M_{\pi}^2}{\mu^2}\,.
 \end{eqnarray}
The role of $\alpha$ and $\beta$ in this equation is that of observable
quantities to be extracted from the data by fitting the two-loop
formula~(\ref{eq:akmsf}), with the $\lambda_i$ fixed by the Roy equations.
The standard prediction for $\alpha$ and $\beta$ is very close to~1 \cite{l1l2}.
On the other hand the contribution from the l.e.c.'s figuring in Eq.~(\ref{eq:xgor1l}) is unknown, at present. The $\rho_i$'s come
from the $O(p^3)$ ${\mathcal L}_{(0,3)}$ Lagrangian, with no derivatives and
three powers of the scalar-pseudoscalar source, the constants $a_i$, $b_i$
and $c_i$ come from the $O(p^4)$ ${\mathcal L}_{(2,2)}$ Lagrangian and the constants
$e_i$, $f_i$ from the $O(p^4)$ ${\mathcal L}_{(0,4)}$ Lagrangian, with analogous
notations.
In principle these l.e.c.'s  should be determined from independent
phenomenological information but, due to the large number of independent
operators, this program seems hopeless.
However we do know the scale dependence of all these l.e.c.'s.
One possibility is thus to treat them as randomly distributed
around zero with an error assigned to them according to naive dimensional
analysis estimates, 
\begin{equation} \label{eq:naive}
\rho_i \sim \pm \frac{1}{\Lambda_{\mathrm{H}}},\quad a_i, b_i, c_i, e_i, f_i \sim \pm
\frac{1}{\Lambda_{\mathrm{H}}^2},
\end{equation}
where the hadronic scale $\Lambda_{\mathrm{H}}$ will be fixed to 1~GeV.
The light quark mass $\hat m$ is written as $\hat m = m_s /r$, where the
quark mass ratio $r$ is strongly correlated to the two-flavor GOR ratio (see
e.g. Fig.~1 of Ref.~\cite{para}) and  $m_s$ is taken conservatively to be
$\sim200$~MeV.
As a consequence of the smallness of the light quark mass, these
contributions will in practice be very small. On the contrary, the
logarithmic terms in Eq.~(\ref{eq:xgor1l}) will be
quite important, specially in the region where $\alpha + 2 \beta$ is
large\footnote{
Notice that the logarithm is absent in the standard case, in which
$\alpha =\beta=1$ in higher orders. Indeed $\alpha$ and
$\beta$ are scale-independent quantities. Since all unknown l.e.c.'s in
Eq.~(\ref{eq:xgor1l}) are at least $O(p^6)$ in the standard counting, there
is no possibility for a standard chiral logarithm at order $p^4$.}.
A further source of uncertainty is therefore the scale $\mu$ inside the
logarithm, i.e. the 
scale at which the estimates~(\ref{eq:naive}) are supposed to hold.
We will take $\mu= M_{\eta} \pm $250~MeV.
The resulting curve is shown as the dashed line in Fig.~\ref{fig:xgor},
the band representing the theoretical uncertainty obtained as explained
above. The main contribution to the error band is represented by the scale
variation inside the logarithm. Due to the large contribution of the
latter it becomes necessary to test the convergence of the chiral series, by
studying the importance of higher orders.
In the following we will compute the $O(p^6)$ double logarithmic corrections
to this result. 
 
\indent

{\bf 4.}~ 
Double chiral logarithms
are among the potentially most dangerous contributions at order $O(p^6)$
\cite{gilbertoll}, because of the smallness of the pion mass. As
first pointed out in Ref.~\cite{bellucci} (see also Ref.~\cite{llogs}) they can be obtained from a 1-loop
calculation, using the fact that, in the renormalization procedure,
non-local divergences must cancel. We set the space-time dimension
$d=4+\omega$ to regulate the theory. The coupling constant $F^2$, which
multiplies the kinetic term for the pions, has dimension $[F^2]=d-2$.
This constant appears as a common factor in the definition of the
generalized chiral Lagrangian. Therefore, a generic l.e.c., with $k$
derivatives and  $l$  powers of the scalar-pseudoscalar source, $c_{(k,l)}$,
has mass dimension $2-k-l$, and does not depend on $\omega$.
We thus replace $F^2$ with $\mu^{ \omega} F^2$, 
making appear explicitly the scale parameter $\mu$ brought in by the 
regularization procedure.         
Since each loop involves a factor $F^{-2}$, the chiral expansion of a
generic amplitude ${\mathcal A}$ up to two loops, apart from an overall
dimensional factor, takes the form,       
\begin{equation} \label{eq:genampl}
 {\mathcal A} \sim {\mathcal A}_{\mathrm{tree}} + \left(
\frac{M_\pi}{\mu}\right)^{\omega} \sum_i P^{(1)}_i (c_a)
g_i^{\mathrm{1-loop}}  + \left(\frac{M_\pi}{\mu} \right)^{2 \omega} \sum_i
P^{(2)}_i (c_a) f_i^{\mathrm{2-loop}},   
\end{equation}
where $P_i^{(1,2)}$ are polynomials in the quark mass and l.e.c.'s,
generically denoted by $c_a$, and $g_i$ and $f_i$ are loop-functions of the
kinematical variables, expressed in terms of dimensionless quantities with
the appropriate inverse powers of $M_{\pi}$. The loop functions are singular
as  $\omega \to 0$,   
\begin{equation} \label{eq:loopfcns}
f_i=\frac{f_{i,2}}{\omega^2} + \frac{f_{i,1}}{\omega} + f_{i,0} + O(\omega),
\quad 
g_i=\frac{g_{i,1}}{\omega} + g_{i,0} +  O(\omega).
\end{equation}
While the one-loop divergences are canceled by the renormalization of the
l.e.c.'s of ${\mathcal L}^{(2)} + {\mathcal L}^{(3)} + {\mathcal L}^{(4)}$,
appearing in ${\mathcal A}_{\mathrm{tree}}$,
\begin{equation} \label{eq:constren}
c_a=\frac{\delta_a}{\omega} + c_a^r + O(\omega),
\end{equation}
the two-loop divergences require the introduction of higher order
local counterterms.
We have defined, compared to the notations of Eq.~(\ref{eq:lecren}),
\begin{equation}
\delta_a = \mu^\omega \frac{\Gamma_a}{16 \pi^2},
\end{equation}
which is actually independent of $\omega$, due to the fact that the
$\beta$-function coefficients $\Gamma_a$ are all proportional to $F^{-2}$
(cfr. Eq.~(\ref{eq:p2ren}) and Ref.~\cite{su2lagr} for the ${\mathcal
L}^{(3)}$ and ${\mathcal L}^{(4)}$ l.e.c.'s).
Notice also that the divergent term in Eq.~(\ref{eq:constren})  always
arises at chiral order $p^4$ in the  generalized counting, as already
mentioned, even if the constant itself is of lower order.
Therefore, since the 1-loop polynomials $P_i^{(1)}$ are already of order
$p^4$, we never 
have to deal, up to and including order $p^6$, with products of the type
$\delta_i \delta_j  g^{\mathrm{1-loop}}$: the polynomials $P_i^{(1)}$ after
the substitution~(\ref{eq:constren}), can be rewritten as
\begin{equation}
P_i^{(1)}(c_a) = P_i^{(1)}(c_a^r) +  \frac{\partial P_i^{(1)}}{\partial c_b} (c_a^r)
\frac{\delta_b}{\omega}.
\end{equation}
For the same reason, since the 2-loop functions are of order $p^6$,
the l.e.c.'s inside the polynomials $P_i^{(2)}$ can be simply replaced by the
corresponding renormalized ones,
\begin{equation}
P_i^{(2)}(c_a) = P_i^{(2)}(c_a^r).
\end{equation}
With the replacements~(\ref{eq:loopfcns}) and~(\ref{eq:constren}) the
$\omega$-dependence of Eq.~(\ref{eq:genampl}) is explicit.
General theorems of renormalization theory require the residues of the
poles in $\omega$ to be polynomials in external momenta and  masses. Thus  the
non-local  divergences, of the type $(\log M_{\pi}) / \omega$ must cancel in
the final result.
This condition amounts to a relation between the coefficients $f_{i,2}$ and
$g_{i,1}$, namely
\begin{equation}
\sum_i \frac{\partial P_i^{(1)}}{\partial c_b}  g_{i,1} \delta_b + 2 \sum_i
P_i^{(2)} f_{i,2} =0.
\end{equation}
This is the reason why the double logarithms can be obtained by a 1-loop
calculation: they only enter through the residues $g_{i,1}$ of the 1-loop
functions.
After renormalization the generic amplitude ${\mathcal A}$ becomes,
\begin{eqnarray}
{\mathcal A} &=& {\mathcal A}_{\mathrm{tree}}^{r} + \sum_i \biggl\{ g_{i,1} \left[
P_i^{(1)}(c_a^r) - \frac{1}{2} \sum_b \frac{\partial P_i^{(1)}}{\partial
c_b}(c_a^r)
\delta_b \log \left(\frac{M_{\pi}}{\mu}\right) \right] \log
\left(\frac{M_{\pi}}{\mu} \right) \nonumber
\\
&&+ g_{i,0} \left[ P_i^{(1)} (c_a^r)+ \sum_b \frac{\partial P_i^{(1)}}{\partial
c_b}(c_a^r) \delta_b \log \left(\frac{M_{\pi}}{\mu} \right)\right] \biggr\}
\nonumber \\
&& + \sum_i
\biggl\{ P_i^{(2)}(c_a^r) \left[ f_{i,0} + 2 f_{i,1} \log \left(\frac{M_{\pi}}{\mu}
\right)
\right] \biggr\}
\end{eqnarray}
Thus we have to compute, besides the 1-loop graphs with the $O(p^2)$
Lagrangian, all the 1-loop graphs with 1 insertion of
operators from the $O(p^4)$ Lagrangian and up to 2 insertions of operators
from the $O(p^3)$ Lagrangian.
At the end, each occurrence of $c_a^r \log (M_{\pi}/\mu)$ must be replaced by
the combination
\begin{equation} \label{eq:replac}
c_a^r \log \left(\frac{M_{\pi}^2}{\mu^2} \right) \rightarrow \left[c_a^r   -
\frac{1}{4} \delta_a \log \left(\frac{M_{\pi}^2}{\mu^2} \right) \right] \log
\left(\frac{M_{\pi}^2}{\mu^2} \right),
\end{equation}
and all contributions beyond $O(p^4)$, but the pure double logarithms, must be
discarded.
Obviously this is only part of the $O(p^6)$ contribution.
Let us examine in more detail what we are considering and what we are
neglecting by this procedure.
The double logarithms are of chiral order $p^6$ and always arise paired
with  single logarithms multiplied by a l.e.c., as in Eq.~(\ref{eq:replac}).
They are always of equal or higher order, compared to the
accompanying single logarithms: the latters, in the generalized counting, can
be of order $p^4$, $p^5$ or $p^6$. We are keeping obviously the $p^4$ single
logarithms, but neglecting the $p^5$ and $p^6$. In any case, we could only
consider them as error sources, since most of the ${\mathcal L}^{(3)}$ and
${\mathcal L}^{(4)}$ l.e.c.'s are unknown experimentally. The fact that, at $O(p^4)$,
the error due to the unknown constants is much smaller than the one coming
from the variation of the scale in the logarithms, is one argument in
favor of keeping only the pure double logarithms.
On the other hand, at least in
 the standard case, the pure double logarithmic contribution to $\alpha$
is by a factor 10 larger than the contributions of the type $l_i L$
\cite{l1l2}.
We are only interested in the relationship between the GOR ratio and the
parameters $\alpha$ and $\beta$.
Since the parameter $\alpha$ is the most correlated to the quark condensate,
we can  expect that the latter also (and therefore the GOR ratio) be
dominated, at $O(p^6)$, by the pure double logarithms.

\indent 

{\bf 5.}~
The first step for computing the $\pi\pi$ amplitude is the
calculation of the axial-axial two-point function, from which we can extract
$F_{\pi}$ and $M_{\pi}$. We display the result for these two quantities,
where all l.e.c.'s, here  and in the following, are renormalized at a scale
$\mu$, but we drop the superscript $^r$ for simplicity:   
\begin{eqnarray}
\frac{F_\pi^2}{F^2} M_{\pi}^2 &=& 2 B \hat m + 4 A \hat m^2 
 + \left( 9 \rho_1 + \rho_2 + 20 \rho_4 + 2 \rho_5 \right) \hat
m^3  \nonumber \\ 
&& + \left( 16 e_1 + 4 e_2 + 32 f_1 + 40 f_2  + 8 f_3 + 96 f_4
\right) \hat m^4  \nonumber \\
&& + 4 a_3 M_{\pi}^2 \hat m^2 -
 \frac{M_{\pi}^2}{32 \pi^2 F_{\pi}^2}  \left( 3 M_{\pi}^2 + 20 A \hat m^2
 \right) \log \frac{M_{\pi}^2}{\mu^2} \nonumber \\
&& + \left[ \frac{33}{8} +\frac{65}{2} \frac{A \hat m^2}{M_{\pi}^2} + 60
\left( \frac{A \hat m^2}{M_{\pi}^2} \right)^2 \right]  M_{\pi}^2 \left(
\frac{M_\pi^2}{16 
\pi^2 F_\pi^2} \log \frac{M_{\pi}^2}{\mu^2}
\right)^2  , \\
F_{\pi}^2 &=& F^2 \left[ 1 + 2 \xi^{(2)} \hat m + \left( 2 a_1 + a_2 +
4 a_3 + 2 b_1 - 2 b_2 \right) \hat m^2 \right. \nonumber \\
&&  \left. - \frac{M_{\pi}^2}{8 \pi^2
F_{\pi}^2} \log \frac{M_{\pi}^2}{\mu^2}
 + \frac{7}{2}
\frac{M_{\pi}^4}{F_{\pi}^4} \left( \frac{1}{16 \pi^2} \log
\frac{M_{\pi}^2}{\mu^2} 
\right)^2   \right]. 
 \end{eqnarray}
While the double chiral logarithm for $F_{\pi}$ is the same as in the
standard case, for $M_{\pi}$ there are additional double logarithmic
contributions which would be relegated by the standard counting at orders
$p^8$ and $p^{10}$. Whether these additional corrections are important or not,
depends on the ratio $A \hat m^2/ M_{\pi}^2$, i.e. on the deviation of the
GOR ratio from 1. The $\pi\pi$ amplitude can be brought to the form of
Eq.~(\ref{eq:akmsf}), explicitly displayed in Ref.~\cite{pipi1}, with the
following expressions for the parameters $\alpha$, $\beta$, $\lambda_1,\ldots
,\lambda_4$ in terms of the l.e.c.'s\footnote{These results were formerly
reported in Ref.~\cite{frascati}.}:
\begin{eqnarray}
\alpha &=& \frac{F^2}{F_\pi^2 M_\pi^2} \biggl\{2 B \hat m + 16 A \hat m^2 -
4 M_{\pi}^2 \xi^{(2)} \hat m  
%\nonumber \\
%&& 
+ \left( 81 \rho_1 + \rho_2 + 164 \rho_4 + 2 \rho_5 \right) \hat
m^3  \nonumber \\
&& - 8 M_{\pi}^2 \left( 2 b_1 - 2 b_2 - a_3 - 4 c_1 \right) \hat m^2 
\nonumber \\
&& + 16 \left( 6 A a_3 + 16 e_1 + e_2 + 32 f_1 + 34 f_2 + 2 f_3 +
72 f_4 \right) \hat m^4  \nonumber \\
&& - \frac{M_{\pi}^2}{32 \pi^2 F_{\pi}^2} \left( 4 M_{\pi}^2 + 204 A \hat m^2  + 528 \frac{ A^2 \hat
m^4}{M_{\pi}^2} \right) \log \frac{M_{\pi}^2}{\mu^2}  \nonumber \\
&& - \frac{ 1}{32 \pi^2 F_{\pi}^2} \left[ M_{\pi}^4 + 88 A \hat m^2
M_{\pi}^2 + 528 A^2 \hat m^4 \right] \nonumber \\
&& + M_{\pi}^2 \left[ \frac{533}{72}  + \frac{18817}{30} \frac{A \hat
m^2}{M_{\pi}^2} +
\frac{61076}{15} \left( \frac{A \hat m ^2}{M_{\pi}^2} \right)^2 \right.
\nonumber \\
&& \left. + 5808
\left( \frac{A \hat m ^2}{M_{\pi}^2} \right)^3 \right] \left(
\frac{M_{\pi}^2}{16 \pi^2 F_{\pi}^2} 
\log \frac{M_{\pi}^2}{\mu^2} \right)^2 \biggr\}, \\
\beta &=& 1 + 2 \xi^{(2)} \hat m - 4 {\xi^{(2)}}^2 \hat m^2 + 2 \left( 3 a_2
+ 2 a_3 + 4 b_1 + 2 b_2 + 4 c_1 \right) \hat m^2  \nonumber  \\
&& - \frac{4 M_{\pi}^2}{32 \pi^2 F_{\pi}^2} \left( 1 + 10 \frac{ A \hat
m^2}{M_{\pi}^2 } \right) \left( \log \frac{M_{\pi}^2}{\mu^2} + 1 \right)
\nonumber \\
&& + \left( \frac{151}{36} M_{\pi}^4 + \frac{400}{3} M_{\pi}^2 A \hat m^2 +
420 A^2 \hat m^4 \right) \left[  \frac{1}{16 \pi^2 F_{\pi}^2}
\log \frac{M_{\pi}^2}{\mu^2} \right]^2 , \\
\lambda_1 &=& 2  l_1 - \frac{1}{48 \pi^2} \left( \log
\frac{M_{\pi}^2}{\mu^2}  + \frac{ 4}{ 3} \right)
+ \left( \frac{25}{18} + \frac{130}{9} \frac{A \hat m^2}{M_{\pi}^2}
\right) \left[  \frac{M_{\pi}^2}{16 \pi^2 F_{\pi}^2}
\log \frac{M_{\pi}^2}{\mu^2} \right]^2 , \\
\lambda_2 &=&  l_2 - \frac{1}{48 \pi^2} \left(\log
\frac{M_{\pi}^2}{\mu^2}  + \frac{ 5}{ 6} \right) 
+ \left( \frac{5}{3} + \frac{80}{9} \frac{A \hat m^2}{M_{\pi}^2} \right)
\left[  \frac{M_{\pi}^2}{16 \pi^2 F_{\pi}^2}
\log \frac{M_{\pi}^2}{\mu^2} \right]^2 , \\
\lambda_3 &=& \frac{10}{9} \left[ \frac{1}{16 \pi^2} \log
\frac{M_{\pi}^2}{\mu^2} \right]^2, \hspace{1cm} \lambda_4 = -\frac{5}{18}
\left[ \frac{1}{16 \pi^2} \log 
\frac{M_{\pi}^2}{\mu^2} \right]^2.
\end{eqnarray}
It is easy to check that these formulae, when restricted to the standard
case,  agree with the ones displayed in Ref.\cite{l1l2} based on the
complete two-loop calculation \cite{loropipi}.  

Eliminating the constant $A$ in favor of $\alpha$ and  the ${\mathcal
L}_{(2,1)}$ coupling $\xi^{(2)}$ in favor
of $\beta$, one can express the GOR ratio of Eq.~(\ref{eq:gor}) as a function
of the combination $\alpha + 2 \beta$,   
\begin{eqnarray} \label{eq:xgorll}
x_2^{\mathrm{GOR}} && =  \left.x_2^{\mathrm{GOR}} \right|_{\mathrm{1-loop}}
+\left[ \frac{11}{60} - 
\frac{4169}{1080} (\alpha + 2 \beta)  + \frac{5639}{1620} (\alpha + 2
\beta)^2 \right. \nonumber \\
&& \left.
- \frac{121}{108} (\alpha + 2 \beta)^3 \right] \left( \frac{M_{\pi}^2}{16 \pi^2 F_{\pi}^2} \log
\frac{M_{\pi}^2}{\mu^2} \right)^2   + O \left( M_\pi^2 \hat m \log
M_{\pi}^2, M_\pi^2 \hat m \right),
\end{eqnarray}
where $x_2^{\mathrm{GOR}} |_{\mathrm{1-loop}}$ is the $O(p^4)$ result
corresponding to Eq.~(\ref{eq:xgor1l}).
We are neglecting single-logarithmic and constant contributions which start
at order $p^5$ in the generalized counting (cfr. discussion at the end of
the previous section).
The function~(\ref{eq:xgorll}) is shown in Fig.~\ref{fig:xgor} as the full line (lower
curve), together with the theoretical uncertainty, represented by the
shaded band and  estimated  varying the
ChPT renormalization scale $\mu$ by $\pm 250$~MeV around the value
$\mu=M_{\eta}$, with the unknown l.e.c.'s considered as error sources, as in
Eq.~(\ref{eq:naive}). In the lower band also, the error comes predominantly
from the variation of the scale. 

\indent

{\bf 6.}~As it is clear from Fig.~\ref{fig:xgor}, the contribution of the
double logarithms to
$x_2^{\mathrm{GOR}}$ is quite large, specially in the ``extreme''
generalized case of large $\alpha + 2 \beta$, and tends to compensate the
$O(p^4)$ single logarithms. Still it is small enough in order to maintain
the validity of the chiral expansion, the $O(p^6)$ weighting about a half of
the $O(p^4)$ corrections. Moreover, the scale dependence is almost
completely canceled by the double logarithms (cfr. the width  of the shaded
bands around the two curves). We stress that, at $O(p^4)$ the scale-dependence
arises because of our ignorance about the l.e.c.'s. It reflects the
uncertainty about the scale at which the estimates~(\ref{eq:naive}) are
supposed to hold. Indeed the effective theory, up to a given order of
accuracy, defines a perfectly consistent theory, in the sense
that the scale dependence from the chiral logarithms is compensated by the
scale dependence of the l.e.c.'s [see Eq.~(\ref{eq:lecbeta})].
The non-renormalizability of the theory means that, as we want to increase
the accuracy, we are forced to consider more and more l.e.c.'s, which
implies a loss of predictive power of the theory.
Therefore, if we knew  the values
of the l.e.c.'s at some scale, the complete $O(p^4)$ result would be
scale-independent. On the contrary, an $O(p^6)$ calculation in the double
logarithmic approximation introduces a scale dependence, which would be
absorbed by the remaining $O(p^6)$ corrections, including further unknown
l.e.c.'s. It is remarkable that these two distinct sources of
scale-dependence almost completely cancel with each other.
This is certainly a signal of reduced theoretical uncertainty, although one
must admit that the cancellation is fortuitous.
A complete two-loop calculation with the generalized Lagrangian, would not
be very useful, due to the flourishing of unknown l.e.c.'s. Double chiral
logarithms are only part of this full two-loop calculation, but their
contribution is known and unambiguous, apart from the scale-dependence.
Moreover, there are reasons to believe that for some observables, namely the
S-wave, isoscalar scattering length, or the parameter $\alpha$, they
constitute -at least in the standard case, where a complete $O(p^6)$
calculation is available- the bulk of the $O(p^6)$ corrections
\cite{gilbertoll}.
 We remind that the status of $\alpha + 2 \beta$ in this
relationship is that of an observable quantity, extracted from the data using
the full {\em two-loop} six-parametrical formula~(\ref{eq:akmsf}), with the
constants $\lambda_i(\alpha,\beta)$ determined {\em e.g.} from the recent solution of
the Roy equations \cite{roysol}. This procedure 
yields, through Eq.~(\ref{eq:xgorll}), an experimental value for the
two-flavor GOR ratio, with a theoretical uncertainty that we estimate to be
twice the width of the lower band of Fig.~{\ref{fig:xgor}.
It is worth recalling that a GChPT fit to the old Rosselet data 
\cite{rosselet} gives $\alpha= 2.16 \pm  0.86$, $\beta=1.074 \pm 0.053$
\cite{pipi1}, where the error was dominated by the statistics, and no
correlations between the two constants where taken into account.
The experimental uncertainty will certainly be reduced by the new data from
the $K_{e4}$ experiment, E865 at Brookhaven \cite{truol}, with tenfold
statistics compared to the Cern-Munich one \cite{rosselet}. A further
increase in statistics is also expected in the planned NA48-2 $K_{e4}$
experiment at CERN, which will provide a very welcome complementary set of
data in  terms of different systematic uncertainties \cite{na48}.  

\indent

{\bf Acknowledgements.}~ This work was partly done at IFAE, Universitat
Autonoma de Barcelona, under EEC-TMR program, Contract N. CT980169
(EURODA$\Phi$NE).

\end{document}